\documentclass[11pt]{article}
\usepackage[a4paper, margin={26.5mm}]{geometry}

\usepackage[utf8]{inputenc}
\usepackage[T1]{fontenc}
\usepackage{amsmath}
\usepackage{amssymb}
\usepackage{amsthm}
\usepackage{mathtools}
\usepackage{bbm}
\usepackage{authblk}
\usepackage{braket}
\usepackage{parskip}
\usepackage[hidelinks]{hyperref}
\hypersetup{
  breaklinks=true,
  pdftitle={Uncertainty Relations Relative to Phase-Space Quantum Reference Frames},
  pdfauthor={Miguel Jorquera Riera and Leon Loveridge},
}

\newtheoremstyle{mystyle}
{.5em}{0em}        
{\addtolength{\leftskip}{1em}\addtolength{\rightskip}{1em}}        
{}                
{\bfseries}       
{:}               
{ }               
{}                

\theoremstyle{mystyle}

\newtheorem{proposition}{Proposition}
\newtheorem{theorem}{Theorem}
\newtheorem{example}{Example}
\newtheorem{definition}{Definition}

\usepackage[
  backend=biber,
  sorting=none,
  doi=false,
  url=false,
  giveninits=true,
  maxbibnames=99,
]{biblatex}

\DeclareFieldFormat*{title}{\mkbibemph
    {\iffieldundef{doi} {\iffieldundef{eprint} {\iffieldundef{url}{#1}
    {\href{\strfield{url}}{#1}}}
    {\href{https://arxiv.org/abs/\strfield{eprint}}{#1}}}
    {\href{https://doi.org/\strfield{doi}}{#1}}}}
\DeclareFieldFormat*{journaltitle}{#1}
\AtEveryBibitem{\clearfield{issn}}
\newcommand{\id}{\mathbbm 1}
\newcommand{\Y}{\yen}
\newcommand{\bor}{{\cal B}}
\newcommand{\hr}{\H_\R}
\newcommand{\hs}{\H_\S}

\newcommand{\tr}{{\rm Tr}}
\newcommand{\ketbra}[2]{\ket{#1}\!\bra{#2}}
\renewcommand{\iint}{\int\hspace{-2.5mm}\int}

\newcommand{\A}{{\sf A}}
\newcommand{\B}{{\sf B}}
\newcommand{\C}{\Gamma}

\newcommand{\D}{\Delta}
\newcommand{\E}{{\sf E}}
\newcommand{\G}{{\sf G}}
\renewcommand{\H}{{\cal H}}
\newcommand{\K}{{\cal K}}
\newcommand{\la}{\lambda}
\newcommand{\M}{{\sf M}}
\renewcommand{\P}{{\sf P}}
\newcommand{\Q}{{\sf Q}}
\newcommand{\R}{{\cal R}}
\newcommand{\rr}{\mathbb R}
\newcommand{\s}{\sigma}
\renewcommand{\S}{{\cal S}}
\newcommand{\w}{\omega}

\begin{document}

\title{Uncertainty Relations Relative to\\ Phase-Space Quantum Reference Frames}
\date{}

\author[1]{Miguel {Jorquera Riera}\thanks{\href{mailto:miquel.jorquera.riera@fmph.uniba.sk}{miquel.jorquera.riera@fmph.uniba.sk}}}
\affil[1]{\footnotesize RCQI, Institute of Physics, Slovak Academy of Sciences, Dúbravská cesta 9, Bratislava 84511, Slovakia }

\author[2]{Leon Loveridge\thanks{\href{mailto:leon.d.loveridge@usn.no}{leon.d.loveridge@usn.no}}}
\affil[2]{Quantum Technology Group, Department of Science and Industry Systems, University of South-Eastern Norway, 3616 Kongsberg, Norway}

\maketitle

\begin{abstract}
\noindent We study Heisenberg's uncertainty relation relative to a quantum reference frame (QRF). 
We introduce the QRF as a covariant phase-space observable, show that when described relative to it, position and momentum appear compatible, and derive novel, frame-relative uncertainty relations.
This is achieved by constructing a joint observable for position and momentum, and calculating the variances of its margins.
We then verify that in the classical limit of the QRF, the standard uncertainty relations are recovered, fortifying claims that standard quantum theory must be understood relative to an external, classical frame. These results may open up new research directions at the interface between QRFs and incompatibility.
\end{abstract}

\section{Introduction}

A fundamental aspect of our description of the physical world is that certain basic properties, such as the position or momentum of an object, must be specified relative to a reference system, or frame.
In a world composed of only quantum systems, such reference bodies are called quantum reference frames (QRFs). Crucially, the explicit incorporation of QRFs into the physical description allows for quantum mechanical observables to attain a form invariant under a given symmetry group. With this observation in mind, we revisit one of the cornerstones of quantum theory: the Heisenberg uncertainty relation ${\D(Q)\D(P)\ge 1/2}$ \cite{heisenberg1927anschaulichen} (we work throughout in units in which ${\hbar = 1}$). At first sight, this relation either does not reflect Galilei invariance, since it is built on the `absolute', rather than relative, position and momentum/velocity, or is not consistent with the universality of quantum theory, relying on some external, classical frame. We resolve this tension by introducing a QRF for position and momentum jointly, thereby restoring invariance when system and frame have equal mass, and derive new uncertainty relations which take into account the presence of the frame. 

A general goal in the study of QRFs is to understand how to describe quantum states, observables, and measurements relative to a QRF. A new wave of research was initiated in \cite{giacomini2019quantum} and generalized in \cite{de2020quantum}, dedicated to understanding how to transform between `perspectives' given by different frames, with interesting results arising when the quantum property of superpositions of frame states is considered. In the present paper, another quantum property of the frame is investigated: the \emph{incompatibility} of position and momentum. We follow the operational framework introduced in \cite{loveridge2012quantum,Miyadera2015e,Loveridge2017a} and developed further in e.g. \cite{carette2023,JGthesis,fewster2024quantum}, described below. We find that relative to a joint frame for position and momentum, these quantities `appear' compatible, and we derive novel frame-relative uncertainty relations, which in a classical limit of the QRF, recover the original relations, reinforcing the view that standard quantum mechanics is drawn up relative to a \emph{classical} external frame. Our work contrasts other work on QRFs, phase space, and uncertainty relations (e.g. \cite{aharonov1984quantum,dickson2004view,de2021perspective,lake2023quantum}) in the use of covariant phase-space observables, and the resulting interplay between frame-relative and standard uncertainty relations, which has not been addressed to date. [{\bf Note}: Since our paper first came out, a treatment by Suleymanov, Carmi and Cohen of (among other things) the frame-dependence of uncertainty relations given in the perspective-neutral framework appeared \cite{suleymanov2025relativity}. Comparing the approach given here with theirs is work for the future.]

\section{Background}
\subsection{Notation}

Given a complex Hilbert space $\H$ with bounded operators $B(\H)$, an \textit{observable} is a normalized positive operator-valued measure (POVM) ${\E:\bor(\Sigma)\to B(\H)}$ (e.g. \cite{busch1995operational}) acting on the Borel $\s$-algebra $\bor(\Sigma)$ of a topological space $\Sigma$, understood as representing the outcomes of some experiment.

The standard description in which observables are represented by self-adjoint operators is recovered via the spectral theorem when ${\Sigma = \rr}$ and each $\E(X)$ is a projection ($\E$ is then called a spectral measure). Such an observable will be called \textit{sharp}, and all others \textit{unsharp}. We will refer to both POVMs and self-adjoint operators as observables, as context demands.

The state space $\S(\H)$ is given by the positive trace-class operators of unit trace, with the pure states as the rank-$1$ projections (or occasionally unit vectors that define them). The pairing of a state $\rho$ and an observable $\E$ gives rise to a probability measure $\la^\E_\rho$ on $\Sigma$, given through the Born rule trace expression
\begin{equation}\label{eq:Born1}
    \la^\E_\rho(X) := \tr[\rho \E(X)].
\end{equation}
If $\E$ (on $\rr$) is sharp, its expectation value in a state $\rho$ takes the form ${\left \langle \E \right \rangle_\rho = \int_\rr x d\la^\E_\rho(x)}$. We will be interested in the sharp position and momentum observables---respectively $Q$ (with spectral measure ${\Q:\bor(\rr)\to B(L^2(\rr))}$) and $P$ (spectral measure $\P$), acting in (a dense domain of) $L^2(\rr)$. We freely oscillate between mathematicians' and physicists' notation, e.g.
\begin{equation}
Q=\int_\rr \!x\,d\Q(x)=\int_\rr \!x\ketbra{x}{x}dx,
\end{equation}
with the spectral measure as usual obtained through $\Q(X)= \chi_X(Q) = \int_X\, d\Q(x)= \int_X \ketbra{x}{x}dx$.

\subsection{Quantum Reference Frames}

QRFs are quantum systems, relative to which other systems may be described. The subject has a long history (see, e.g., \cite{wick1952intrinsic,aharonov1967charge,wick1970superselection,aharonov1984quantum} for the early work and \cite{Bartlett2007} for a review of the state of the field, from an information-theoretic viewpoint, up to 2007), and various frameworks now exist, catering to different needs and physical situations; for example, the perspective-neutral approach \cite{Vanrietvelde:2018pgb,de2021perspective} most naturally arises in the study of constrained systems. 

The point of departure here is the principle that `true' observables are invariant under relevant symmetry transformations; QRFs may be used to construct invariants as described shortly. Following \cite{carette2023}:
\begin{definition} Let $\cdot : G \times \Sigma \to \Sigma$ denote a continuous transitive left action. A \textit{quantum reference frame} is a system of covariance ${\R\coloneqq(\hr,U_\R,\E_\R)}$, where $\hr$ is a Hilbert space carrying a (strongly continuous, projective) unitary representation $U_\R$ of $G$, and the POVM ${\E_\R: \bor(\Sigma)\to B(\hr)}$ is covariant:
\begin{equation}\label{eq:cov}
    U_\R(g)\E_\R(X)U_\R(g)^* = \E_\R(g\cdot X).
\end{equation}
\end{definition}

There are various classes of QRFs; if $U_\R$ is the left regular representation in $L^2(G)$ and ${\E_\R(X)f=\chi_X f}$, where $\chi_X$ is the set indicator function, this is an example of an \emph{ideal} frame, so named so as to reflect various desirable properties. When $G$ is identified with $\Sigma$, the frame is called \emph{principal}. Here, the main frames of interest will be principal, but not ideal.

\begin{example}\label{ex:Q}
For the group of spatial translations, $G=(\rr,+)$,
we may fix a QRF ${(L^2(\rr), U_\R,\Q_\R)}$. Here, $\Q_\R$ is the position observable of the reference system, and the unitary representation is ${U_\R(q)=e^{-iqP_\R}}$, where $q\in\rr$ and the translation generator $P_\R$ is the momentum operator. The covariance condition now reads
\begin{equation}
e^{-iqP_\R}\Q_\R(X)e^{iqP_\R} = \Q_\R(X+q).
\end{equation}
or ${e^{-iqP_\R}Q_\R e^{iqP_\R} = Q_\R-q \id_\R}$ for the corresponding self-adjoint operator.
\end{example}

A quantum system $\S$ combined with a QRF $\R$ is described by the Hilbert space ${\hs \otimes \hr}$, with (projective) unitary representation ${U\coloneqq U_\S \otimes U_\R:G \to B(\hs \otimes \hr)}$. We may obtain \emph{relativized} observables, which are invariant under $U$ (by conjugation), as follows:

\begin{definition}
Let ${(\hr,U_\R,\E_\R)}$ be a principal QRF for $G$. The \textit{relativization map} ${\Y^{\E_\R}:B(\hs) \to B(\hs \otimes \hr)}$ is given by
\begin{equation}
    \Y^{\E_\R}(A)=\int_G U_\S(g)AU_\S(g)^* \otimes d\E_\R(g).
\end{equation}
\end{definition}

POVMs can also be relativized via ${(\Y^{\E_\R}\circ\E_\S)(X)=\Y^{\E_\R}(\E_\S(X))}$. 
It is readily verified that ${U(g)\Y^{\E_\R}(A)U(g)^* = \Y^{\E_\R}(A)}$ for all ${g\in G}$, i.e., relativization yields an observable that is invariant under the given representation of $G$. Relativization $\Y^{\E_\R}$ is a quantum channel (a unital completely positive map) (see e.g., \cite{Loveridge2017a} for further properties, \cite{glowacki2023quantum,fewster2024quantum} for generalizations to some homogeneous spaces of $G$ and \cite{glowacki2024relativization} for a categorical analysis). 
 \begin{example}\label{ex:Qrel}
Using the QRF from Example \ref{ex:Q}, the relativization map applied to the spectral measure $\Q_\S$ of the system's position operator $Q_\S$ yields 
\begin{equation}
    \Y^{\Q_\R} (\Q_\S(X)) = \iint_{\rr^2}\chi_X(x-y)d\Q_{\S}(x) \otimes d\Q_{\R}(y).
\end{equation}
This is the spectral measure of relative position ${Q^{\rm rel}\coloneqq Q_\S \otimes \id_\R - \id_\S \otimes Q_\R}$, which can be seen by noting that ${\chi_X(Q^{\rm rel}) = \Q^{\rm rel}(X)}$, and writing 
\begin{subequations}\begin{align}
Q^{\rm rel}
&=\iint_{\rr^2}(x-y)d\Q_{\S}(x) \otimes d\Q_{\R}(y)\\
&=\iint_{\rr^2} (x-y)\ketbra{x,y}{x,y}\, dx\,dy,
\end{align}\end{subequations}

where ${\ket{x,y}\coloneqq\ket x_\S\otimes\ket y_\R}$. Therefore, we may write 
\begin{equation}
    \Y^{\Q_\R} (Q_\S) = Q_\S \otimes \id_\R - \id_\S \otimes Q_\R =Q^{\rm rel}.
\end{equation}
 \end{example}
The relative momentum ${P_\S \otimes \id_\R - \id_\S \otimes P_\R}$ is obtained by choosing the frame observable to be the spectral measure of momentum, which is covariant under momentum translations. We note, however, that this is a \emph{different} representation of $\rr$, given by ${V_\R(p)=e^{ipQ_\R}}$, under which $Q_\R$ is invariant. The purpose of this paper is to construct a QRF that is covariant under translations in both position and momentum, to which we will soon turn.

It is also important to describe the system relative to the frame prepared in a state $\w$; this is done via the \emph{restriction map} ${\C_\w: B(\hs \otimes \hr) \to B(\hs)}$, defined such that
\begin{equation}
    \tr[\C_{\w }(\Xi)\rho] = \tr[(\rho \otimes \w)\Xi],
\end{equation}
where the above holds for all system states $\rho$ and any ${\Xi \in B(\hs \otimes \hr)}$. For each state $\w$ of the frame, $\C_{\w }$ is a quantum channel, known as a \textit{conditional expectation} \cite{kadison2004non}, and gives a description of the system contingent on the frame preparation $\w$; here the frame is viewed as being externalized.

Relativization combined with restriction gives rise to the \textit{frame-conditioning map} 
${\Y^{\E_\R}_\w:B(\hs)\to B(\hs)}$, given by
\begin{subequations}\begin{align}\label{eqs:conditioning1}
    \Y^{\E_\R}_\w\coloneqq &\ \C_{\w }\circ \Y^{\E_\R},\\
    \Y^{\E_\R}_\w(A) = & \int_G U_\S(g)A\,U_\S(g)^* d\la_{\w}^{\E_\R}(g),\label{eqs:conditioning2}
\end{align}\end{subequations}
where the measure $\la_{\w}^{\E_\R}$ on $G$ is the Born rule measure \eqref{eq:Born1}, i.e., ${\la_{\w}^{\E_\R}(X)= \tr [\w \E_\R(X) ]}$. The frame-conditioning map therefore realizes an arbitrary system observable relative to the given frame as an invariant, and then externalizes the frame, given a particular frame state. 

\begin{example}\label{ex:fcsm1}
Choose the frame as in Examples \ref{ex:Q}, \ref{ex:Qrel}, with ${\w=\ketbra{\psi}{\psi}_\R}$. We find
\begin{equation}\begin{aligned}
&\Y^{\Q_\R}_\psi(\Q_\S(X))= \int_\rr\Q_\S(X+y)\,d \la^{\Q_\R}_\psi(y)\\
&=\int_\rr\Q_\S(X+y)\,|\psi(y)|^2\,dy = (\la^{\Q_\R}_\psi \!\star \Q_\S) (X),
\end{aligned}\end{equation}
where `$\star$' denotes measure convolution. Equivalently, the above expression may be written as
${\Y^{\Q_\R}_\psi(\Q_\S(X)) = (\chi_X \star |\psi|^2)(Q_{\S})}$. Since $|\psi|^2$ is never a delta for any vector $\psi \in L^2(\rr)$, the frame-conditioned position is strictly never equal to the sharp position of the system, but can be made arbitrarily close by taking the frame's wavefunction $\psi$ to be very tightly peaked around $x=0$.
 \end{example}

This highlights a general observation: The tightness of the localization of $\la_{\w}^{\E_\R}$ around the identity of $G$ dictates the quality of approximation of $\Y^{\E_\R}_\w(A)$ by $A$ \cite{Loveridge2017a,Miyadera2015e,carette2023}, with arbitrarily good approximation being achieved when $\la_{\w}^{\E_\R}$ is close to a delta at the identity. $\Y^{\E_\R}(A)$ is understood as the `true', relative and invariant quantity, and the frame-conditioned observable $\Y^{\E_\R}_\w(A)$ is understood as representing such a quantity conditioned on frame preparation $\w$. 
 
It has been shown in \cite{Loveridge2017a,carette2023} for ideal frames that all observables can be recovered for a sufficiently localized frame state in the above sense (actually, one can weaken the ideality somewhat, and use POVMs with the \emph{norm}-$1$ property \cite{heinonen2003norm}). The fact that the description in terms of the system alone has had such empirical success is likely due to the ubiquity of `good' quantum frames; indeed, this may be an essential feature of how we experience the `classical world'. The uncertainty relation prohibits good localization for position and momentum together, and this observation will form the basis of the frame-relative uncertainty bounds derived later.

The structure that emerged above---the convolution of an operator measure with a scalar measure or density---is a generic means by which to add noise to an observable, as made precise in the following definition. 
\begin{definition}\label{def:smea1}
Let ${\mu:\bor(\rr)\to[0,1]}$ be an absolutely continuous (probability) measure on $\rr$. A \textit{smearing} $\E^\mu$ of the observable ${\E:\bor(\rr)\to B(\H)}$ by $\mu$ is an observable given by 
\begin{equation}\label{eq:ftwss}
  \E^\mu(X) \coloneqq (\mu \star \E)(X) =\int_\rr \mu (x+X)d\E(x)
\end{equation}
\end{definition}
The absolute continuity means we may write ${\mu(X)=\int_X e(x)dx}$ for some density $e$, as in Ex. \ref{ex:fcsm1} with $e = |\psi|^2$. Our main interest is in smeared position and momentum observables $\Q^\mu,\P^\nu$, which are approximations of $\Q,\P$ in the sense that they are covariant (under position and momentum translations) and have finite error relative to $\Q,\P$ \cite{busch2007universal}.

\subsection{Covariant phase-space observables}
As emphasized in \cite{Busch2007}, Heisenberg's classic result---usually viewed in the negative as a limitation on preparation and measurement---can be understood in a positive light as delineating possibilities of joint measurements of position and momentum, provided appropriate approximations are made. This point can be made systematically through the use of POVMs and a more general, operational notion of compatibility than is afforded by the usual statement on commuting self-adjoint operators. 
\begin{definition}
    Let $\A$ and $\B$ be POVMs on $\Omega_A$ and $\Omega_B$ respectively. $\A$ and $\B$ are called \textit{compatible/jointly measurable} if there is a POVM $\M$ on $\Omega_A \times \Omega_B$ which has $\A$ and $\B$ as marginals, i.e., ${\A(X)=\M(X \times \Omega_B)}$ and ${\B(Y)=\M(\Omega_A\times Y)}$ for all measurable subsets $X\subseteq\Omega_A$ and $Y\subseteq\Omega_B$. Else, they are \textit{incompatible}.
\end{definition}
Equivalently, $\A$ and $\B$ are compatible if and only if they have joint probability distributions in all states. 
Joint measurability of $\A$ and $\B$ is equivalent to their commutativity (i.e., ${[\A(X),\B(Y)]=0}$ for all $X,Y$) when at least one of them is sharp. Position and momentum are of course incompatible, but by adding noise through smearing as in Def. \ref{def:smea1}, it is possible to obtain jointly measurable unsharp approximators $\Q^\mu$, $\P^\nu$.

The question of how much smearing is needed to make $\Q^\mu$ and $\P^\nu$ compatible is answered with \emph{covariant phase-space observables} (e.g. \cite{Busch2016a}).
Writing ${U(q)\coloneqq e^{-iqP}}$, which effects a spatial translation by $q \in \rr$, and ${V(p)\coloneqq e^{ipQ}}$ for the translations by momentum ${p\in \rr}$, the \emph{Weyl operators} are defined by ${W(q,p)\coloneqq e^{i\frac{qp}2}U(q)V(p)}$. $W$ is an irreducible projective unitary representation of $\rr^2$ acting in $L^2(\rr)$, with multiplication ${W(q,p)W(q',p')= e^{\frac{-i(qp'-q'p)}{2}}W(q+q',p+p')}$.
A covariant phase-space observable ${\G^T:\bor(\rr^2) \to B(L^2(\rr))}$ is then defined as
\begin{equation}\label{eq:cpso}
    \G^T(Z)\coloneqq \frac{1}{2\pi}\int_Z W(q,p)TW(q,p)^*dq\,dp,
\end{equation}
where $T$ is a density operator acting in $L^2(\rr)$. For instance, $T$ may be the vacuum of a single mode optical field, in which case the integrand is a set of coherent states labeled by $q$ and $p$ \cite{Perelomov} (see \cite{de2021perspective} for a treatment of this case in the perspective-neutral approach). For any $T$, such an observable is indeed covariant under phase-space translations:
\begin{equation}
    W(q,p)\G^T(Z)W(q,p)^* = \G^T(Z + (q,p)).
\end{equation} 
Any covariant phase-space observable $\G$ is of the form $\G^T$ for some unique state $T$ \cite{Busch2016a, holevo1979, holevo1982probabilistic, werner1984quantum}. The margins of $\G^T$ are smeared position and momentum observables
\begin{subequations}\begin{alignat}{3}
    &\G^T(X \times \rr) &&= (\mu_T \star \Q)(X) &&=\Q^{\mu_T}(X) ;\\
    &\G^T(\rr \times Y) &&=(\nu_T \star \P)(Y)&&= \P^{\nu_T}(Y),
\end{alignat}\end{subequations}
where
\begin{subequations}\label{eqs:smearing}
\begin{alignat}{2}\label{eqs:smearing1}
    &\mu_T(X) &&= \tr[\Pi T\Pi\, \Q(X)];\\
    &\nu_T(Y) &&= \tr[\Pi T\Pi\, \P(Y)].\label{eqs:smearing2}
\end{alignat}\end{subequations}
 Here, $\Pi$ is the parity operator, i.e., ${(\Pi f)(x) = f(-x)}$ for ${f \in L^2(\rr)}$. The following fact \cite{Busch2007,werner2004uncertainty,carmeli2004position} is of fundamental importance:
\begin{theorem}
\label{th:ibc}
 $\Q^\mu$ and $\P^\nu$ are jointly measurable if and only if there is a $T$ such that $\mu = \mu_T$ and $\nu = \nu_T$ as in Eqs. \eqref{eqs:smearing}.
\end{theorem}
Hence, smeared position and momentum observables are compatible exactly when there is a state $T$ such that their smearing measures are of the form \eqref{eqs:smearing}.
$\G^T$ therefore dictates which smeared position and momentum may be observed in an approximate joint measurement. We note that since the phase-space observable $\G^T$ is completely determined by $T$, we view $T$ not as a particular state preparation of the system being described, but as an intrinsic attribute of the measuring apparatus pertaining to $\G^T$.

\section{Quantum Reference Frames in phase space}

We now let the (necessarily non-ideal) QRF be defined by a covariant phase-space observable $\G_\R^{T'}$ (we will typically use the `prime' for the frame). For any ${A\in B(\hs)}$, the relativized observable
\begin{equation}
    \Y^{T'}(A)\coloneqq\iint_{\rr^2} W_\S(q,p)A\,W_\S(q,p)^* \otimes d\G_\R^{T'}(q,p)
\end{equation}
is invariant under translations in phase space. We have used the shorthand ${\Y^{T'}\coloneqq\Y^{\G^{T'}_\R}}$ to indicate the choice of covariant phase-space observable appearing in the integral. We note that momentum translation invariance would only be expected under Galilei symmetry in the case of system and frame having equal mass, and hence if imposing Galilei symmetry, such a stipulation is required.

\begin{proposition}
  Let $\G_\S^T$ be a covariant phase-space observable in $\H_\S$.
  For ${X,Y\in\bor(\rr)}$, the margins of ${\Y^{T'}\circ\G_\S^T}$ give rise to smeared relative position and momentum observables:\begin{subequations}\label{eq:doc}
\begin{align}
(\Y^{T'}\circ\G_\S^T)(X\times\rr)&=
(\mu_T \star \mu_{T'}\star \Q^{\rm rel})(X)\nonumber\\&=
\label{eq:doc1}
\Y^{T'}(\Q_\S^{\mu_T}(X));\\
(\Y^{T'}\circ\G_\S^T)(\rr \times Y)&= (\nu_T \star \nu_{T'}\star\P^{\rm rel})(Y)\nonumber\\&=
\label{eq:doc2}
\Y^{T'}(\P_\S^{\nu_T}(Y)).
\end{align}\end{subequations}
\end{proposition}
The quality of the approximation is dictated by the spread of the smearing measures; this can be captured by the standard deviation $\D$ or the variance $\D^2$ (though other measures of spread are better---e.g. the overall width; see e.g. \cite{Busch2016a}). For instance, for any state $\s$, 
\begin{equation}
    \D^2(\mu_T\star \mu_{T'} \star \Q^{\rm rel}, \s) = \D^2(\mu_T) + \D^2( \mu_{T'}) + \D^2( \Q^{\rm rel}, \s).
\end{equation}
The observables in Eqs. \eqref{eq:doc} are always unsharp, but e.g. the smeared relative position can be made arbitrarily close to the sharp relative position by taking $T$ and $T'$ in \eqref{eq:doc1} highly peaked in position (or for momentum in \eqref{eq:doc2}). We may compose the relativization with the restriction, yielding frame-conditioned observables
\begin{equation}
    \Y^{T'}_\w(A)=\iint_{\rr^2} W_\S(q,p)AW_\S(q,p)^* d\la_{\w}^{\G_\R^{T'}}(q,p),
\end{equation}
where we use again the shorthand ${\Y^T_\w\coloneqq\Y^{\G^T_\R}_\w}$ and recall that the integration measure is obtained through the Born rule ${\la_\w^{\G_\R^{T'}}(Z) = \tr{[\w\G_\R^{T'}(Z)]}}$ for any reference state $\w$ and measurable ${Z \subset \rr^2}$.
For the sharp position and momentum, this gives
\begin{subequations}\label{eq:qprel}\begin{alignat}{2}
\label{eq:qrel}
    &(\Y^{T'}_\w \circ \Q_\S)(X) &&= (\la_{\w,T'}^{\Q_\R} \star \Q_\S) (X);\\
    \label{eq:prel}
    &(\Y^{T'}_\w \circ \P_\S)(Y) &&= (\la_{\w,T'}^{\P_\R} \star \P_\S) (Y),
\end{alignat}\end{subequations}
where the measure \begin{subequations}\label{eq:titrr}\begin{alignat}{3}
 &\la_{\w,T'}^{\Q_\R}(X)&&\coloneqq\tr[\w\,\Q_\R^{\mu_{T'}}(X)]&&=(\mu_{T'}\star\la^{\Q_\R}_\w)(X);\\
&\la_{\w,T'}^{\P_\R}(Y)&&\coloneqq\tr[\w\,\P_\R^{\nu_{T'}}(Y)]&&=(\nu_{T'}\star\la^{\P_\R}_\w)(Y),
\end{alignat}
\end{subequations}
Therefore, the frame-conditioned sharp position (and momentum) is never sharp, and the quality of approximation depends again on the frame preparation $\w$, and on $T'$. The restrictions of the observables given in Eqs. \eqref{eq:doc} yield 
\begin{subequations}\label{eq:urr}\begin{align}
(\Y^{T'}_\w \circ \G_\S^T)(X\times\rr)
&=(\la_{\w,T'}^{\Q_\R} \star\mu_T \star\Q_\S)(X)\nonumber\\&=
\label{eq:urr1}
\Y^{T'}_\w(\Q_\S^{\mu_{T}}(X));\\
(\Y^{T'}_\w \circ\G_\S^T)( \rr\times Y)&=(\la_{\w,T'}^{\P_\R} \star \nu_T \star\P_\S)(Y)\nonumber\\&=
\label{eq:urr2}
\Y^{T'}_\w(\P_\S^{\nu_{T}}(Y)).
\end{align}\end{subequations}
Thus, the margins of a frame-conditioned covariant phase-space observable are smeared positions and momenta, coinciding with the frame-conditioned compatible smeared positions and momenta. 
The contributions to the (in)accuracy of the approximation to their sharp counterparts come from $T$ and $T'$---internal features of the system and frame respectively---and from the frame preparation $\w$. 

\section{Breaking incompatibility of position and momentum}
It is well known (e.g. \cite{heinosaari2015incompatibility}), that given a set ${I=\{\A_1,...,\A_n\}}$ of compatible observables in $B(\H)$, and any channel ${\Lambda:B(\H) \to B(\K)}$, the observables in $\Lambda (I)$ are also compatible. On the other hand, if some observables in $I$ are incompatible, the question whether those in $\Lambda (I)$ are compatible is nontrivial and motivates the following definition \cite{heinosaari2015incompatibility}:

\begin{definition}
    Let $I$ be a set of observables as above, not all of which are compatible, and let $\Lambda$ be a channel. Then $\Lambda$ is \textit{incompatibility-breaking} for $I$ if all the observables in $\Lambda(I)$ are compatible.
\end{definition}

We apply this to $I={\{\Q_\S,\P_\S\}}$ and ${\Lambda =\Y^{T'}_\w: B(\hs) \to B(\hs)}$ (recall that ${\hr=\hs=L^2(\rr)}$):

\begin{theorem}
    For any states ${\w,T'\in\S(\hr)}$, the channel $\Y^{T'}_\w$ breaks the incompatibility of ${\{\Q_\S,\P_\S\}}$.
\end{theorem}

To prove the claim, we need to demonstrate the existence of a joint observable that has ${\Y^{T'}_\w \circ \Q_\S}$ and ${\Y^{T'}_\w \circ \P_\S}$ as marginals. This can be done by direct construction: set
\begin{equation}
    \M^{T',\w}_\S\coloneqq (\la^{\Q_\R}_\w\times \la^{\P_\R}_\w)\star\G^{T'}_\S.
\end{equation}
This indeed has margins \eqref{eq:qprel}:
\begin{subequations}\begin{alignat}{7}
&\M^{T',\w}_\S(X\times\rr)&&=(\la^{\Q_\R}_{\w,T'}\star\Q_\S)(X)&&=(\Y^{T'}_\w \circ \Q_\S)(X);\\
&\M^{T',\w}_\S(\rr\times Y)&&=(\la^{\P_\R}_{\w,T'}\star\P_\S)(Y)&&=(\Y^{T'}_\w \circ \P_\S)(Y).
\end{alignat}\end{subequations}

Thus, relative to a covariant phase-space frame, the incompatible $\Q_\S$ and $\P_\S$ `appear' compatible, as might be expected, since position and momentum cannot be defined more sharply than the frame with respect to which they are described (see also \cite{Loveridge2017a}). 

\section{Uncertainty Relations}
The uncertainty relation 
\begin{equation}\label{eq:shur}
    \D(\Q,\rho)\D(\P,\rho) \geq 1/2
\end{equation}
for the standard deviation of position and momentum of a system in state $\rho$ is one of the most iconic in (quantum) physics, taken to express the incompatibility of position and momentum as sharp observables. However, the inequality \eqref{eq:shur} does not include a QRF, and therefore should be reconsidered using the ideas developed thus far.

The uncertainty principle is typically understood to contain two conceptually distinct impossibility statements: that expressed in \eqref{eq:shur}, which demonstrates that no state can be prepared that is jointly sharply localized with respect to $Q$ and $P$, and secondly that $Q$ and $P$ cannot be jointly measured. Thus \eqref{eq:shur} may be called a \emph{preparation uncertainty relation}. Now, noting that 
a compatible smeared pair ${\Q^{\mu_T}=\mu_T\star\Q}$ and ${\P^{\nu_T}=\nu_T\star\P}$ obey a more restrictive bound \cite{Busch2007} than their sharp counterparts in \eqref{eq:shur}, namely 
\begin{equation}\label{eq:sur}
    \D(\Q^{\mu_T},\rho)\D(\P^{\nu_T},\rho)\geq 1,
\end{equation}
leads to the subject of \emph{measurement uncertainty relations} \cite{busch2013disturbance,busch2014measurement,busch2014colloquium,Busch2016a,bullock2018measurement}. That is, the degree of approximation error that must be accommodated in order to allow for a joint measurement, which in this case can be given by the trade-off relation on the standard deviations of the smearing measures:
\begin{equation}\label{eq:pur}
\D(\mu_T)\D(\nu_T)\ge 1/2.
\end{equation}
In what follows, we explore preparation and measurement uncertainty relative to a quantum reference frame. We use the following, which we state without proof:
\begin{proposition}\label{lemma:UR}
    Let ${a,b,c,d}$ be measures, and ${x,y\ge0}$. 
    If ${\D(a)\D(b)\ge x}$ and ${\D(c)\D(d)\ge y}$, then 
    ${\D(a\star c)\D(b\star d)\ge x+y}$.
\end{proposition}

We now consider frame-conditioned uncertainty relations, all of which are tight. We begin by relativizing the sharp position and momentum observables with respect to a covariant phase-space observable of the frame, yielding
\begin{equation}\label{eq:0001}
    \D(\Y^{T'}_\w\circ\Q_\S, \rho) \D(\Y^{T'}_\w\circ\P_\S, \rho) \ge 3/2.
\end{equation}
Since ${\D^2(\Y^{T'}_\w\!\!\circ\Q_\S, \rho) = \D^2(\mu_{T'}) \!+\! \D^2(\la^{\Q_{\R}}_{\w})\!+\! \D^2(\Q_{\S},\rho)}$, in order to make this quantity small, it is necessary that ${\D(\la^{\Q_{\R}}_{\w}) = \D(\Q_{\R}, \w)}$ is small, in which case ${\D (\P_\R, \w)}$ is large, making explicit the intuition that a frame preparation $\w$ cannot provide a good origin for position and momentum simultaneously. We also observe that the inequality \eqref{eq:0001} comprises both preparation and measurement uncertainty. A version that also incorporates measurement uncertainty for the system is obtained by relativizing a smeared compatible pair $\Q_\S^{\mu_{T}}$ and $\P_\S^{\nu_T}$, giving 
\begin{equation}\label{eq:gur}
\D (\Y^{T'}_\w \circ \Q_\S^{\mu_{T}}, \rho)\D (\Y^{T'}_\w \circ \P_\S^{\nu_{T}}, \rho)\ge 2.
\end{equation}
We note that this inequality corresponds exactly, via Eqs. \eqref{eq:urr}, to the case that one covariant phase-space observable is relativized with respect to another, and then the relevant margins are taken. That the bound is twice that of \eqref{eq:sur} reflects the uncertainty present in the frame. 

The above cases correspond to the setting that the frame observable contains compatible smearings of position and momentum as margins, and therefore those observables could be measured together in a single experiment. This contrasts the case in which the system observables are separately relativized with respect to the frame's sharp position and momentum observables, as in
\begin{equation}\label{eq:0003}
\D(\Y^{\Q_\R}_\w\circ\Q_\S, \rho)
\D(\Y^{\P_\R}_\w\circ\P_\S, \rho)\ge 1,
\end{equation}
or for a compatible smeared pair of the system,

\begin{equation}\label{eq:0002}
        \D(\Y^{\Q_\R}_\w\circ\Q_\S^{\mu_T}, \rho)
        \D(\Y^{\P_\R}_\w\circ\P_\S^{\nu_T}, \rho)\ge 3/2.
\end{equation}
We notice the symmetry here with \eqref{eq:0001}, which captures the idea that the overall indeterminacy is a relational property of system and frame---it is operationally equivalent to describe a smeared, compatible pair of system position and momentum relative to a frame defined by sharp position and momentum observables respectively, or to describe sharp position and momentum observables relative to a frame which contains a compatible smeared pair as marginals. There is, however, a conceptual distinction, in that the two situations represent different experimental arrangements.

\subsection{Classical limit of the QRF}

We have seen that if we stipulate only translation invariance, tight localization of the state of the QRF at $x=0$ allows for the recovery (to arbitrary accuracy) of the standard description of the quantum system in terms of the usual observables and states. In this sense the frame may be viewed as having been externalized, or as is commonly understood, classical. However, in the phase space setting, a choice must be made regarding whether the frame state is well localized in position, or momentum, given the associated uncertainty trade-off---here the reduced description is always `fuzzy' in comparison to the standard setup.

We wish to identify the features of a genuinely classical frame in the phase space setting, which in contrast to the case of only translation invariance, cannot be attained within the formalism of quantum mechanics. We take the view that the hallmark of classical physics is the joint measurability of all observables and the existence of states with zero uncertainty with respect to position and momentum (and all other sharp observables). Therefore, we examine the frame-dependent uncertainty relations in the setting that we declare, by fiat, that the frame state $\w$ is the phase-space point $(0,0)$ (or equivalently, the Dirac measure at $(0,0)$), i.e., is a classical pure state, perfectly localized in phase space. Of course, this is not consistent with quantum mechanics, but we expect that it corresponds to a rigorous classical limit in the small $\hbar$ regime of the reference frame \cite{landsman2017foundations}. 
We write $\w_0$ and $T'_0$ for the classical states localized at $(0,0)$. Then, \eqref{eq:0001} and \eqref{eq:gur} become
\begin{align}\label{eq:tvfb}
\D (\Y^{T'_0}_{\w_0}\circ\Q_\S,\rho)\D (\Y^{T'_0}_{\w_0}\circ\P_\S,\rho)&\ge 1/2;\\ \label{eq:tvfb2}
\D (\Y^{T'_0}_{\w_0} \circ \Q_\S^{\mu_{T}},\rho)\D (\Y^{T'_0}_{\w_0} \circ \P_\S^{\nu_{T}},\rho)&\ge 1,
\end{align}
from which we obtain the bounds given in Eqs. \eqref{eq:shur} and \eqref{eq:sur} respectively. These correspond to the standard uncertainty relations, formulated without a frame. Thus we may conclude that the standard uncertainty relations appear to be drawn up relative to an external, \emph{classical} reference frame. The question of experimentally distinguishing if the bounds \eqref{eq:tvfb}, \eqref{eq:tvfb2}, rather than \eqref{eq:0001}, \eqref{eq:gur} are realized in nature, is open.

\section{Conclusions}
In this paper we have introduced a (non-ideal) quantum reference frame as a covariant phase-space observable, shown that these give rise to an incompatibility-breaking channel for position and momentum, and derived a number of uncertainty relations which depend on both system and frame. Through the classical limit, our observations lend credence to the view that standard quantum mechanics has an external, classical frame in the background---a view widely held but never before demonstrated in this context. Further work includes analyzing frame changes in the sense of \cite{giacomini2019quantum,de2020quantum,Vanrietvelde:2018pgb,Vanrietvelde:2018dit,de2021perspective,castro2021relative} and particularly \cite{carette2023} where informational completeness of covariant phase-space observables \cite{kiukas2012} may play an interesting role.

\section{Acknowledgments}
We express our gratitude to Thomas Galley, Dan McNulty, Anne-Catherine de la Hamette, Anaí Echeverría, Jan G{\l}owacki, and M. Hamed Mohammady for very valuable feedback on early and late drafts, and to Kjetil B{\o}rkje for several helpful discussions. We also thank two anonymous referees for valuable suggestions. M.J.R. acknowledges funding provided by the IMPULZ program of the Slovak Academy of Sciences under the Agreement on the Provision of Funds No. IM-2023-79.

\printbibliography
\end{document}